\documentclass[runningheads]{llncs}
\usepackage{graphicx}
\usepackage{cite} 
\usepackage{enumitem}
\usepackage{siunitx}
\usepackage{ulem}
\usepackage[table,xcdraw]{xcolor}
\usepackage{diagbox}
\usepackage{multirow}
\usepackage{booktabs}
\usepackage{mathrsfs}
\usepackage{amsmath}
\usepackage{amsfonts}
\usepackage{hyperref}
\hypersetup{
    colorlinks=true,
    linkcolor=blue,
    filecolor=magenta,      
    urlcolor=cyan,
    pdftitle={Overleaf Example},
    pdfpagemode=FullScreen,
    }
\begin{document}

\title{Unveiling Fairness Biases in Deep Learning-Based Brain MRI Reconstruction}
\titlerunning{Unveiling Fairness Biases in DL-Based Brain MRI Reconstruction}

%
\author{
Yuning Du\inst{1}\and Yuyang Xue\inst{1}\and Rohan Dharmakumar\inst{2} \and Sotirios A.Tsaftaris \inst{1, 3}}
\index{Du, Yuning \and Xue, Yuyang\and Dharmakumar, Rohan\and Tsaftaris, Sotirios}

\authorrunning{Y.Du et al.}
\institute{School of Engineering, The University of Edinburgh, Edinburgh, EH9 3FG, UK \and Krannert Cardiovascular Research Center, Indiana University School of Medicine, Indianapolis, Indiana, USA\and The Alan Turing Institute, London, NW1 2DB, UK\\}
%
\maketitle              
\begin{abstract}
Deep learning (DL) reconstruction particularly of MRI has led to improvements in image fidelity and reduction of acquisition time. 
In neuroimaging, DL methods can reconstruct high-quality images from undersampled data. 
However, it is essential to consider fairness in DL algorithms, particularly in terms of demographic characteristics. This study presents the first fairness analysis in a DL-based brain MRI reconstruction model. The model utilises the U-Net architecture for image reconstruction and explores the presence and sources of unfairness by implementing baseline Empirical Risk Minimisation (ERM) and rebalancing strategies. Model performance is evaluated using image reconstruction metrics. 
Our findings reveal statistically significant performance biases between the gender and age subgroups. Surprisingly, data imbalance and training discrimination are not the main sources of bias. This analysis provides insights of fairness in DL-based image reconstruction and aims to improve equity in medical AI applications.

\end{abstract}
\section{Introduction}


Magnetic resonance imaging (MRI) is routinely used to help diagnose or ascertain the pathophysiological state in a noninvasive and harmless manner. However, MRI is characterised by long acquisition times. There is an  interest in improving imaging fidelity whilst reducing acquisition time. 
A solution is to subsample the frequency domain (k-space). This introduces aliasing artefacts in the image domain due to the violation of the Nyquist sampling theorem, causing difficulties such as biomarkers extraction and interpretation in neuroimaging. 

Recently, deep learning (DL) methods based on convolutional neural networks (CNNs) have been proposed to reconstruct high-quality images from the undersampled k-space data~\cite{lin2021artificial}. By learning complex patterns from large amounts of training data and filling in missing k-space data, these DL models successfully reconstruct images that closely resemble those obtained through fully sampled acquisitions. Advances in DL-based image reconstruction enable both accelerated acquisition and high-quality imaging, providing significant benefits. 

Deep learning methods may be subject to biases (e.g., from the training dataset) which can lead to fairness and lack of equity. For example, recent studies have shown that image segmentation algorithms can be unfair: Puyol-Ant{\'o}n et al.~\cite{puyol2022fairness} found racial bias can exist in DL-based cine CMR segmentation models when training with a race-imbalanced dataset. This leads us to ask: \textit{Could DL-based image reconstruction algorithms be also unfair?} 

To date, such a, at least empirical, study is lacking, and this article precisely addresses this gap. 
Our primary objective is to investigate the biases in the algorithm resulting from demographic information present in the training data. To the best of our knowledge, this is the first fairness analysis in a DL-based image reconstruction model. We make the following contributions:
\begin{enumerate}
    \item[$\bullet$]We identify existing bias in performance between gender and age groups using the publicly available OASIS  dataset~\cite{marcus2007open}. 
    \item[$\bullet$]We investigate the origin of these biases by mitigating imbalances in the training set and training paradigm with different bias mitigation strategies. 
    \item[$\bullet$]We discuss the factors that may impact the fairness of the algorithm, including inherent characteristics and spurious correlations. 
\end{enumerate}

\section{Background} 

\subsection{Fairness Definitions}
Amongst the various definitions of fairness,since we study the fairness for different demographic subgroups,
we consider only group fairness in our analysis. 

\noindent \textbf{Group Fairness}: 
Group fairness aims to ensure equitable treatment and outcomes for different demographic or subpopulation groups. It recognises the potential for biases and disparities in healthcare delivery and seeks to address them to promote fairness and equity~\cite{hellman2008discrimination}. To ensure fairness, equalised odds~\cite{hardt2016equality} is used as a criterion that focuses on mitigating bias, stating as ``The predictor $\hat{Y}$ satisfies equalised odds with respect to protected attribute A and outcome Y, if $\hat{Y}$ and A are independent conditional on Y.'' The criterion can be formulated as
\begin{equation}\label{eq1}
\forall y \in\{0,1\}:P(\hat{Y}=1|A=0,Y=y) = P(\hat{Y}=1|A=1,Y=y).
\end{equation}

\noindent \textbf{Fairness in Image Reconstruction}:
It requires the reconstructed image to faithfully represent the original one without distorting or altering its content based on certain attributes such as race, gender, or other protected attributes. 

When applying equalised odds as the fairness criterion, while the original equation focuses on fairness in predictive labels, image reconstruction tasks typically involve matching pixel values or image
representations. Thus, we reformulate the problem based on probabilistic equalised odds, as proposed by~\cite{pleiss2017fairness}. We let $P \subset \mathbb{R}^k$ be the input space of an image reconstruction task, $(\mathbf{x},\mathbf{y}) \sim P$ represent a patient, with $\mathbf{x}$ representing the undersampled image, and $y$ representing the fully sampled image or ground truth image. Also, we assume the presence of two groups $g_1, g_2 \subset P$, which represent the subsets defined by the protected attribute $\mathbf{A}$. Fairness using probabilistic equalised odds is formulated as:
\begin{equation}\label{eq2}
\forall \mathbf{y} \in \mathcal{Y}: \mathbb{E}(\mathbf{x}, \mathbf{y}) \sim g_1 [f(\mathbf{x}) \mid \mathbf{Y}=\mathbf{y}] = \mathbb{E}(\mathbf{x}, \mathbf{y}) \sim g_2 [f(\mathbf{x}) \mid \mathbf{Y}=\mathbf{y}].
\end{equation}
Here, $f$ represents the DL-based reconstruction network. With this formulation, we aim to achieve fairness by ensuring that the quality or fidelity of the reconstructed image is consistent across different data distributions irrespective of different demographic characteristics. 

\subsection{Source of Bias}
\noindent \textbf{Data imbalance} can be a significant source of bias in medical scenarios~\cite{zong2022medfair}. It can refer to the imbalanced distribution of demographic characteristics, such as gender and ethnicity, within the dataset. For example, the cardiac magnetic resonance imaging dataset provided by the UK Biobank~\cite{raisi2021cardiovascular} is unbalanced with respect to race: $> 80\%$ of the subjects are of white ethnicity, resulting in unequal representation of features correlated to ethnicity. This imbalance can introduce bias in the analysis and interpretation of the data. 

\noindent \textbf{Training discrimination}  is another source of bias, possibly occurring concurrently with data imbalance~\cite{kamiran2012data}. An imbalanced dataset can lead to imbalanced minibatches drawn for training. Hence, the model mainly learns features from the dominant subgroup in each batch, perpetuating bias in the training process. 

\noindent \textbf{Spurious correlations}  can also contribute to bias~\cite{zong2022medfair}. This refers to the presence of misleading or incorrect correlations between the training data and the features learned by the model. For instance, a model can learn how to classify skin diseases by observing markings made by dermatologists near lesions, rather than fully learning the diseases~\cite{winkler2019association}. This is particularly likely to happen in the minority subgroup due to limited presence in training dataset, leading to overfitting during the training process and further exacerbating bias. 

\noindent \textbf{Inherent characteristics}  can also play a role in bias, even when the model is trained with a balanced dataset~\cite{zong2022medfair}. Certain characteristics may inherently affect the performance of different subgroups. For instance, in skin dermatology images, lesions are often more challenging to recognise in darker skin due to lower contrast compared to lighter skin. As a result, bias based on ethnicity can still exist even if the dataset is well-balanced in terms of proportions. 
 
\section{Methods}
\label{sec:Methods}
\textbf{Our main goal}: Our goal is to identify the bias in image reconstruction models and any potential sources of bias related to demographic characteristics. 
To investigate fairness in image reconstruction tasks, we systematically design and conduct experiments that eliminate potential origins of bias w.r.t. various demographic characteristics. 
We start by establishing a baseline model using Empirical Risk Minimisation (ERM) to assess the presence of bias in relation to diverse demographic subgroups. Then, we employ a subgroup rebalancing strategy with a balanced dataset in terms of demographic attributes, to test the hypothesis that bias is caused by data imbalance. Then, we use the minibatch rebalancing strategy to evaluate the effects of  training discrimination for each subgroup.

\noindent \textbf{Reconstruction Networks}:  
We use a U-Net~\cite{ronneberger2015u} as the backbone for the reconstruction network. 
The reconstruction network is trained using undersampled MRI brain scans, which are simulated by applying a random Cartesian mask to the fully sampled k-space data. Details of the data and the experimental setup of the reconstruction network are provided in Section~\ref{sec:Experimental Analysis}. 

\noindent \textbf{Baseline Network}:
We follow the principle of Empirical Risk Minimisation (ERM)~\cite{vapnik1991principles}. ERM seeks to minimise the overall risk of a model by considering the entire population, instead of the composition of specific groups and hence without controlling for the distribution of protected attributes.  

\noindent \textbf{Subgroup Rebalancing Strategy}:
This strategy aims to examine the performance 
when a perfectly balanced dataset of the protected attributes is used. Instead of randomly selecting data from the entire dataset to define a training set, the training set consists of an equal number of subjects from different subgroups according to demographic characteristics. This approach ensures that all subgroups have equal chances during the training phase, helping us identify if data imbalance is the source of bias. 

\noindent \textbf{Minibatch Rebalancing Strategy}:
This strategy examines the performance 
when balanced minibatches in terms of protected attributes are used to eliminate discrepancy before training~\cite{puyol2021fairness}. Hence, 
each minibatch has an equal presence of subjects with different demographic characteristics and all subgroups have an equal opportunity during each iteration to influence the model weights. 

\noindent \textbf{Evaluation Metrics}:
Although several fairness metrics have been proposed, most of the current work is focused on image classification and segmentation tasks, which may not be directly applicable to image reconstruction tasks. Therefore, we analyse the fairness of image reconstruction using image reconstruction metrics and statistical analysis. The performance of the reconstruction is evaluated using Structural Similarity Index (SSIM, higher is better), Peak Signal-to-Noise Ratio (PSNR, higher is better) on the patient level. 

To investigate bias between subgroups with different demographic characteristics, we performed the non-parametric Kruskal-Wallis ANOVA test (as available within OriginPro 2023) to test the omnibus hypothesis that there are differences in subgroups with $p < 0.05$ as the threshold for statistical significance. The test will provide Chi-Square value and p-value as results. Higher Chi-Square values indicate the presence of more significant differences between subgroups. This approach allows us to assess the potential bias in the image reconstruction process specifically instead of relying on fairness metrics designed for other tasks. 
%
%

\section{Experimental Analysis}
\label{sec:Experimental Analysis}
\subsection{Dataset and pre-processing}
\label{sec:Data}
\noindent \textbf{Dataset}:
We select the publicly available Open Access Series of Imaging Studies (OASIS) dataset~\cite{marcus2007open} to evaluate the fairness of the image reconstruction task. The initial data set consists of a cross-sectional collection of 416 subjects(316 subjects is healthy and 100 subjects is clinically diagnosed with very mild to moderate Alzheimer’s disease)and for each subject, three or four individual T1-weighted MRI scans obtained in single imaging sessions are included. 
To simulate clinical practice with uncertainty about patients' conditions, we used an entire dataset consisting of a mix of patients, including both healthy subjects and patients with Alzheimer's disease (AD), without providing explicit labels for their conditions. 
To study the fairness regarding inherent demographic information, we choose gender and age information provided in the dataset as the protected attributes. Since the patients are aged 18 to 96, we categorise the patients into young adult (age below 40), middle-aged adult (ages 40 to 65), and older adult (age above 65) according to the criteria proposed by~\cite{slijepcevic2022explaining}.The statistics of the subgroups are summarised in Table~\ref{table1}. 

\begin{table}[t]
\caption{Statistics of demographic subgroups in OASIS. Patients are categorised into young adult (below 40), middle-aged adult (40 to 65) and older adult (above 65). }\label{table1}
\centering
\begin{tabular}{l|c|cc|ccc}
\hline
\multicolumn{1}{c|}{\textbf{Category}} & \multicolumn{1}{c|}{\textbf{All}} & \multicolumn{1}{c}{\textbf{Female}} & \multicolumn{1}{c|}{\textbf{Male}} & \multicolumn{1}{c}{\textbf{Young}} & \multicolumn{1}{c}{\textbf{Middle-aged}} & \multicolumn{1}{c}{\textbf{Older}} \\
\hline
\multicolumn{1}{c|}{\textbf{Count}}& 416 & 256 & 160 & 156 & 82 & 178 \\
\textbf{Proportion (\%)} & 100.0 & 61.5 & 38.5 & 37.5 & 19.7 & 42.8 \\
\hline
\end{tabular}
\vspace{-1.5em}
\end{table}

According to Table~\ref{table1}, there is a clear imbalance in the distribution of demographic characteristics in the OASIS dataset. In the protected attribute gender, the female is the dominant group with $256$ subjects, while the male is the disadvantaged group with only $ 160 $ subjects. In terms of age, compared to the middle-aged adults group, the young and older adults groups are dominant groups. 

\noindent \textbf{Data Pre-processing}:
To ensure the equal size of dataset for methods in Section~\ref{sec:Methods}, the dataset is firstly categorised into six age-gender subgroups(e.g.,middle-aged female adults) and sampled according to the size of minority subgroups, which is $27$ from middle-aged male adults, to maintain the balance for both age and gender distribution among sampled dataset ($162$ subjects in total). Then, we sampled $5$ subjects from six age-gender subgroups to form test set, which is $30$ subjects in total. For the training and validation set, we sampled the rest $22$ subjects from each age-gender subgroups for the rebalancing and minibatch rebalancing strategies, which is $132$ subjects in total. While for the baseline network, the training and validation set are randomly sampled with a size of $132$ subjects. The train-validation-test splits all follow the proportions of $20:2:5$. For each patient, we select 122 central slices out of 208 slices in one volume.

\subsection{Implementation Details}
We employ a U-Net as backbone. Its first feature map size is $32$, with $4$ pooling cascades, resulting in a total of $7.8 \mathrm{M}$ parameters. We employ the Adam optimiser with a learning rate of $10^{-4}$ with a step-based scheduler with a decay gamma of $0.1$. Both the $\ell_1$ loss and the $\mathrm{SSIM}$ loss were incorporated into our experiments. Models were  trained for 40 epochs with batch size 6. 

5-fold cross validation is used to mitigate sample bias. Our experimental setup uses the PyTorch Lightning framework and we trained on an NVIDIA A100 Tensor Core GPUs. The implementation of our code is inspired by the fastMRI repository.\footnote[1]{https://github.com/facebookresearch/fastMRI/} Our code is publicly available at: \url{https://github.com/ydu0117/ReconFairness}.

\subsection{Results}
\begin{figure}[t]
\begin{center}
\includegraphics[width=0.8\textwidth]{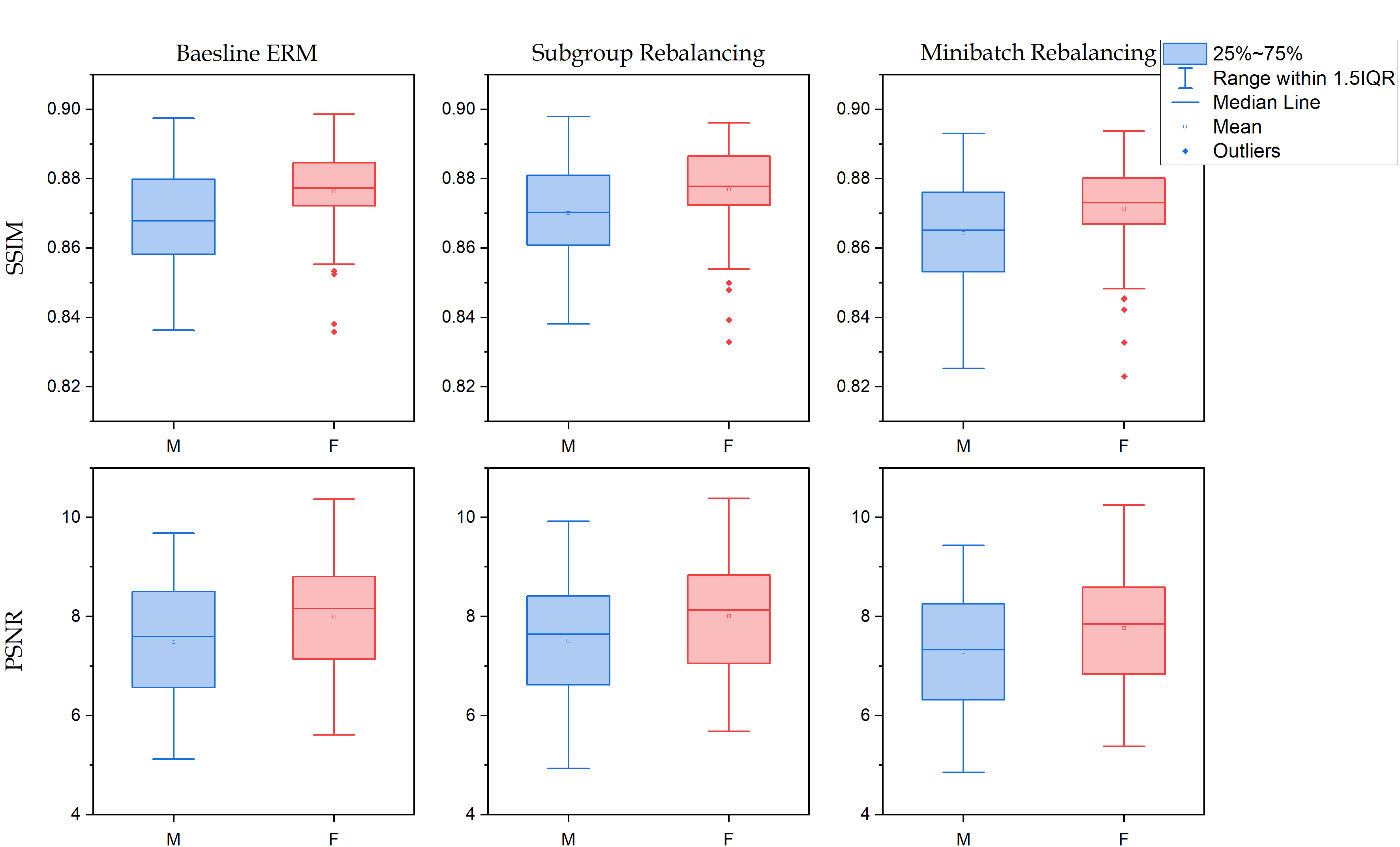}
\end{center}
\caption{Image Reconstruction Performance for Gender Subgroups. In the figure, `F' represents `Female' and `M' represents `Male'. This figure indicates performance gap between two gender subgroups in image reconstruction task under different strategies.} \label{fig1}
\vspace{-1.5em}
\end{figure}

\begin{figure}[t]
\begin{center}
\includegraphics[width=0.8\textwidth]{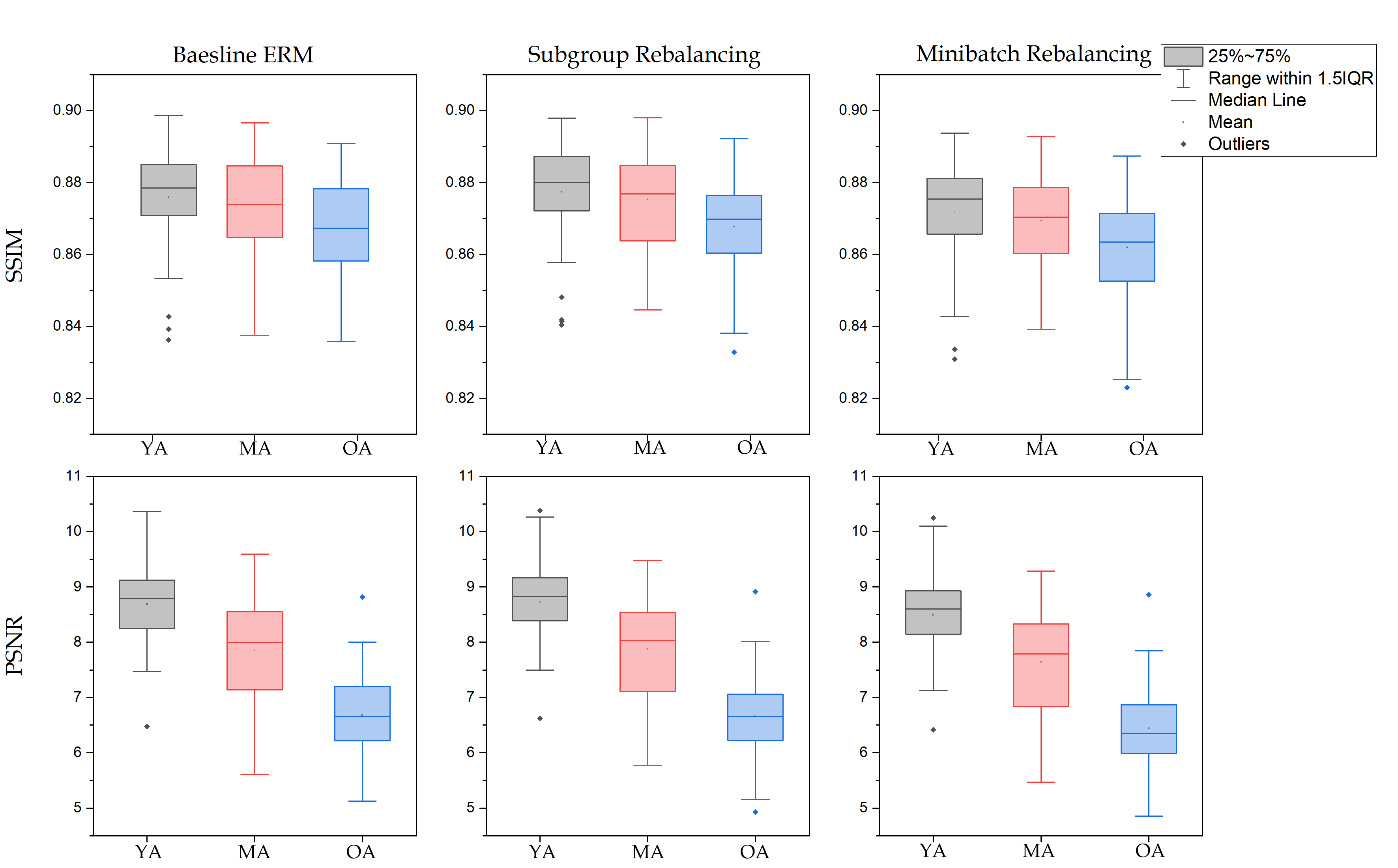}  
\end{center}
\caption{Image Reconstruction Performance for Age Subgroups under Three strategies. In the figure, `YA' represents `Young Adults' , `MA' represents `Middle-aged Adults' and `OA' represents `Older Adults'. This figure indicates performance gaps between age subgroups in image reconstruction task under different strategies.} \label{fig2}
\vspace{-0.5em}
\end{figure}

Table~\ref{table2} reports the SSIM and PSNR results (mean and standard deviation) from 5-fold cross-validation under three different strategies. 
Figures~\ref{fig1} and~\ref{fig2} demonstrate the reconstruction performance of subgroups defined by demographic characteristics. Table~\ref{table3} offers the results of Kruskal-Wallis ANOVA test between demographic subgroups, including p-values and Chi-Square values.  

\begin{table}[t]
\vspace{1em}
    \centering
    \caption{Statistics of Image Reconstruction Performance under three strategies.}
    \resizebox{1.0\linewidth}{!}{%
    \begin{tabular}{c|cc|cc|cc}
        \hline
        & \multicolumn{2}{c|}{\textbf{Baseline ERM}} & \multicolumn{2}{c|}{\textbf{Subgroup Rebalancing}} & \multicolumn{2}{c}{\textbf{Minibatch Rebalancing}} \\
        \cline{2-3} \cline{4-5} \cline{6-7}
        & SSIM & PSNR & SSIM & PSNR & SSIM & PSNR \\
        \hline
         \textbf{Whole} & 0.872 (0.012) & 7.742 (0.112) & 0.867 (0.011) & 7.529 (0.109) & 0.867 (0.011) & 7.529 (0.109) \\
         \hline
        \textbf{Female} & 0.876 (0.010) & 7.999 (0.099) & 0.876 (0.010) & 8.006 (0.095) & 0.871 (0.010) & 7.767 (0.095) \\
        \textbf{Male} & 0.868 (0.013) & 7.485 (0.118) & 0.870 (0.013) & 7.509 (0.117) & 0.864 (0.013) & 7.292 (0.117) \\
        \hline
        \textbf{Young Adults} & 0.876 (0.010) & 8.690 (0.092) & 0.877 (0.009) & 8.729 (0.090) & 0.872 (0.009) & 8.496 (0.090) \\
        \textbf{Middle-aged Adults} & 0.874 (0.011) & 7.859 (0.108) & 0.875 (0.011) & 7.877 (0.106) & 0.869 (0.011) & 7.645 (0.106) \\
        \textbf{Older Adults} & 0.867 (0.010) & 6.676 (0.102) & 0.867 (0.010) & 6.666 (0.099) & 0.861 (0.010) & 6.448 (0.099) \\
        \hline
    \end{tabular}}
    \label{table2}
    \vspace{-1.5em}
\end{table}

\begin{table}[t]
    \centering
    \caption{Kruskal-Wallis ANOVA results for the Three Strategies Testing for Influence of ``Gender'' and ``Age Group''. The results include Chi-Square values with statistical significance (p-value) indicated as  $^{***}\; p < 0.001$, $^{**}\; p < 0.01$, $^{*}\; p < 0.05$.}
    \resizebox{0.7\linewidth}{!}{%
    \begin{tabular}{c|cc|cc}
        \hline
        & \multicolumn{2}{c|}{\textbf{Gender}} & \multicolumn{2}{c}{\textbf{Age Group}} \\
        \cline{2-3} \cline{4-5}
        & SSIM & PSNR & SSIM & PSNR \\
        \hline
        \textbf{Baseline ERM} & $13.44^{***}$ & $6.45^{*}$ & $11.64^{**}$ & $78.91^{***}$ \\
        \textbf{Subgroup Rebalancing} & $10.90^{**}$ & $6.01^{*}$ & $14.64^{**}$ & $81.08^{***}$ \\
        \textbf{Minibatch Rebalancing} & $10.30^{**}$ & $5.44^{*}$ & $13.70^{**}$ & $78.62^{***}$ \\
        \hline
    \end{tabular}}
    \label{table3}
    \vspace{-1.5em}
\end{table}

\noindent \textbf{Presence of Bias}:
Focusing on the baseline ERM model, our results show that there is a significant performance difference between subgroups categorised by gender and age. Among the gender subgroups, the female group outperforms the male group. This difference is obvious in Figure~\ref{fig1}, showing that the baseline model provides better performance for female subjects compared to male subjects. This difference is statistically significant in Table~\ref{table3}. 

Among the three age groups, the results demonstrate an obvious performance gap. Referring to Table~\ref{table2}, the young adults group provides a better performance in all three metrics. 
Furthermore, the results indicate that as age increases, the reconstruction performance worsens (both metrics). The trend is visually evident in Figure~\ref{fig2} and is statistically significant in Table~\ref{table3}. 

\noindent \textbf{Is dataset imbalance the source of unfairness?}
The performance under rebalancing strategies shows that the imbalance of data and the discrimination of training are not the major cause of bias. Specifically, in Table~\ref{table2}, when comparing the performance of the subgroups under different training strategies, the performance gaps evidenced before still exist. 
These biases are also visually illustrated in Figures~\ref{fig1} and~\ref{fig2} and are again statistically significant in Table~\ref{table3}. 


However, the Chi-square values under rebalancing strategies in Table~\ref{table3}, is reduced compared to the baseline ERM network among gender subgroups. 
This reduction in Chi-Square values indicates that the rebalancing either of the training set or the minibatch may mitigate partial bias, illustrating that dataset imbalance and training discrimination towards gender may be sources of bias, but not the main source. However, it is noticeable that the balancing strategies result in performance reduction of the dominant subgroup.

\section{Discussion}

\begin{figure}[t]
\begin{center}
\includegraphics[width=0.6\textwidth]{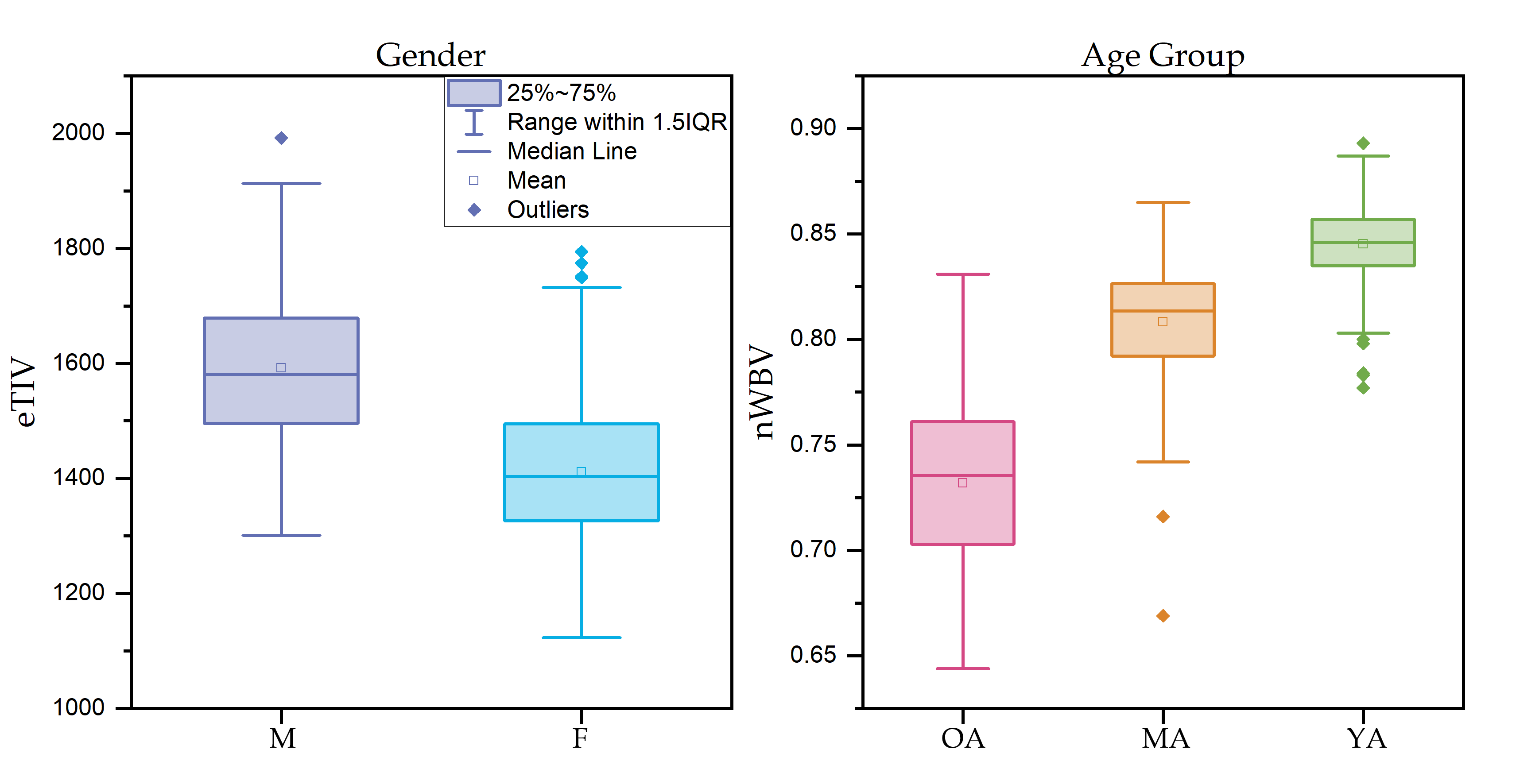}
\end{center}
\caption{Relations between Demographic Features and Neuroanatomy Metrics.} \label{fig3}
\vspace{-1.5em}
\end{figure}

\noindent \textbf{What is the source of unfairness}:
We find that data imbalance and training discrimination do not significantly contribute to bias. Instead, the bias may stem from spurious correlations and inherent characteristics. Specifically, the model may focus on neuroanatomical features that are associated with demographic factors~\cite{ritchie2018sex, gunning2009aging}. In Figure~\ref{fig3}, the relations between demographic features and neuroanatomy metrics including estimated Total Intracranial Volume (eTIV) as well as normalised Whole Brain Volume (nWBV) are analysed. Our results show that women tend to have smaller eTIV compared to men, and young adults have the highest nWBV among age subgroups. Thus, these differences in eTIV between gender and nWBV between age may result in spurious correlations that lead to bias, which requires further investigation in future work. 

\noindent \textbf{Clinical Relevance}:
It is noticeable that the difference in SSIM among subgroups is in the second or third decimal place in some cases. Although the small difference may not be clinically meaningful in practice, it can lead to additional errors and bias in downstream tasks such as segmentation and classification, ultimately leading to inaccurate diagnoses.

\noindent \textbf{Limitations}:
Previous studies~\cite{zhang2022improving} have reported data imbalances among different racial groups due to geographic limitations of the datasets. In our analysis, due to the lack of racial data, the training set may still exhibit an imbalance in terms of race, even if we implement a rebalancing strategy. 

\section{Conclusion} 
In this study, we conducted an initial analysis of fairness in DL-based image reconstruction tasks with respect to demographic characteristics, specifically gender and age. We employed three strategies to investigate the bias caused by these characteristics. Through the use of rebalancing strategies, we found that imbalanced training sets and training discrimination were not the major contributors to bias. However, further investigation is needed to identify the sources of bias in image reconstruction tasks. Correspondingly, we need to propose bias mitigation strategies to ensure fairness in DL-based image reconstruction applications. 

\section*{Acknowledgements}
This work was supported in part by National Institutes of Health (NIH) grant 7R01HL148788-03. Y. Du and Y. Xue thank additional financial support from the School of Engineering, the University of Edinburgh. 
S.A.\ Tsaftaris also acknowledges  the support of Canon Medical and the Royal Academy of Engineering and the Research Chairs and Senior Research Fellowships scheme (grant RCSRF1819\textbackslash 8\textbackslash 25), and the UK’s Engineering and Physical Sciences Research Council (EPSRC) support via grant EP/X017680/1.  The authors would like to thank Dr. Chen and K. Vilouras for inspirational discussions and assistance.
Data used in Sec. \ref{sec:Data} were provided by OASIS-1: Cross-Sectional: Principal Investigators: D. Marcus, R, Buckner, J, Csernansky J. Morris; P50 AG05681, P01 AG03991, P01 AG026276, R01 AG021910, P20 MH071616, U24 RR021382. 
%
\newpage
\bibliographystyle{splncs04} 
\bibliography{reference} 

\end{document}